\documentclass[11pt,showpacs,showkeys,eqsecnum,footinbib,preprint,superscriptaddress]{revtex4}
\usepackage[utf8]{inputenc}
\usepackage[T1]{fontenc}
\usepackage{epsfig}
\usepackage{amsmath}
\usepackage{amsfonts}
\usepackage{amssymb}
\usepackage{color}
\usepackage{graphicx}
\usepackage{url,hyperref}
\usepackage{latexsym} 
\usepackage{longtable}
\usepackage{float}
\usepackage{epstopdf}
\usepackage{url}
\begin{document}
	\title{Scattering states of Dirac particle equation with position dependent mass under the cusp potential}
	\author{M. Chabab}
	\email{mchabab@uca.ma}
	\affiliation{High Energy Physics and Astrophysics Laboratory, Department of Physics, FSSM, Cadi Ayyad University P.O.B. 2390, Marrakesh, Morocco.}
	\author{A. El Batoul}
	\email{elbatoul.abdelwahed@edu.uca.ma}
	\affiliation{High Energy Physics and Astrophysics Laboratory, Department of Physics, FSSM, Cadi Ayyad University P.O.B. 2390, Marrakesh, Morocco.}
	\author{H. Hassanabadi}
	\email{h.hasanabadi@shahroodut.ac.ir}
	\affiliation{Physics Department, Shahrood University of technology, Shahrood, Iran}
	\author{M. Oulne}
	\email{oulne@uca.ma}
	\affiliation{High Energy Physics and Astrophysics Laboratory, Department of Physics, FSSM, Cadi Ayyad University P.O.B. 2390, Marrakesh, Morocco.}
	\author{S. Zare}
	\email{soroushzrg@gmail.com}
	\affiliation{Department of Basic Sciences, Islamic Azad University North Tehran Branch, Tehran, Iran.}
	\date{\today}
	\begin{abstract}
		 We solved the one-dimensional position-dependent mass Dirac equation in the presence of the cusp potential and reported the solutions in terms of the Whittaker functions. We have  derived the reflection and transmission coefficients by making use of  the matching conditions on the wave functions. The effect of position dependent mass on  the reflection and transmission coefficients of the system is duly investigated.
	\end{abstract}
	\maketitle
	\section{Introduction}
 The investigation of scattering and bound states is an important subject in quantum physics.
 In recent years, there has been an increasing interest for exploring nuclear single particle states and  transmission resonance, in the presence of external potentials, through the non relativistic Schrödinger  equation as well as the relativistic Klein Gordon and Dirac equations. Notably, as momentum scattering goes to zero appears in one-dimensional Schr\"{o}dinger equation even potentials, the reflection coefficient goes to unity unless the potential $V (x)$ supports a zero-energy resonance \cite{b1}. In this instance, the transmission coefficient goes to unity, becoming a transmission resonance \cite{b2}. The same result has been recently observed in Dirac equation showing that transmission resonances at low momentum  $(k=0)$ in the particle Dirac takes place \cite{b3} for a potential barrier $V=V(x)$ when the corresponding potential well $ V =-V(x)$ supports a supercritical state. An effort in this direction has been reported in some recent literature\cite{b4,b5,b6,b7,b8,b9,b10}. For example the authors of \cite{b8} have proved that relation between the bound-state energy eigenvalues and transmission resonances  for the Klein-Gordon particle under the Woods-Saxon potential,  which  are the same as what was obtained for the Dirac particle\cite{b10}. On the other hand, solutions of the wave equations have, recently, become interesting in the view of position-dependent mass  formalism. This formalism has been first proposed by Von Roos\cite{b11,b12} in the framework of Schr\"{o}dinger equation.  Also, extensive applications of this formalism have been done in different areas of physics such as condensed matter physics and material science such as electronic properties of semiconductors \cite{b13}, quantum dots \cite{b14}, quantum liquids\cite{b15,b16,b17,b18} , and  atomic nuclei\cite{b19,b20,b21} . In addition, the scattering problem  has been extended, not long ago, to the case where the mass depends on spatially coordinate\cite{b22,b23,b24,b25} .\\
 The cusp potential, despite having a motivating physical structure, has only been the subject of a few researches.  The potential has been investigated within the framework of Dirac equation in \cite{b9,b26,b27}, where the scattering problem, the super-critically condition and the resonant states were well studied. The relativistic Klein Gordon equation under this interaction term was studied in \cite{b28,b30}.  Therefore, our aim in the present work is to investigate the scattering states of Dirac particle equation with position dependent mass under the cusp potential. Particularly, we study the effect of a  position-dependent mass on the  reflection and transmission coefficients. The organisation of this article is as follows. In Section II, we present the theoretical background of the  Dirac equation in the framework of Position dependent mass. In Section III, we discuss the analytical scattering states solutions. Section IV contains the numerical results of the reflection and transmission coefficients.  Finally, our conclusions are drawn in Section V.
\section{Dirac Equation with Position Dependent Mass }
\begin{figure}[H]
	\centering
	\rotatebox{0}{\includegraphics[height=60mm]{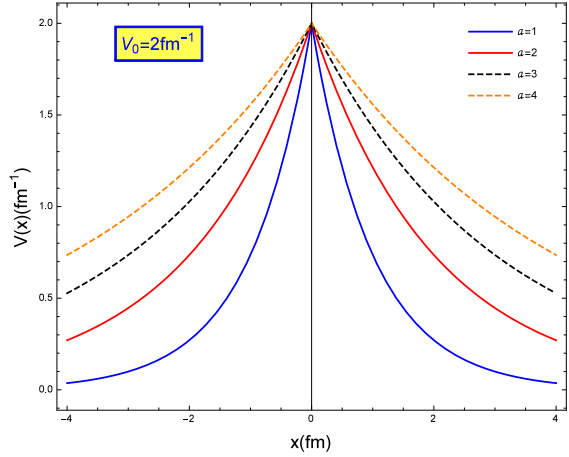}} 
	\caption{Plot of the cusp potential barrier for  $V_0=2fm^{-1}$.}
	\label{fig1}
\end{figure}

In the natural units ($\hbar=c=1$), the relativistic free-particle Dirac equation  is
written as\cite{b31,b32}
\begin{eqnarray}
	\left[i\gamma^{\mu}\partial_{\mu}-m(x)\right]\psi(x)=0,\,
	\label{E1}
\end{eqnarray}
where we assume that the mass of the Dirac particle depends only
on one spatially coordinate $x$. Under the effect of an external
potential $V(x)=eA_0(x)$ (is an electromagnetic potential) and taking the gamma matrices $\gamma_{x}$ and
$\gamma_{0}$ as the Pauli matrices $i\sigma_{x}$ and $\sigma_{z}$,
respectively, the Dirac equation in one-dimension, for a
stationary state  $\psi(x,t)=e^{-iEt}\psi(x)$, becomes
\begin{eqnarray}
	\left\{\left(  \begin{array}{cc}
		0 & 1 \\
		1 & 0 \\
	\end{array} \right)\frac{d}{dx}-[E-V(x)]\left(  \begin{array}{cc}
	1 & 0 \\
	0 & -1 \\
\end{array} \right)+m(x)\left(  \begin{array}{cc}
1 & 0 \\
0 & 1 \\
\end{array} \right)\right\}\left(  \begin{array}{c}
\varphi_{1}(x)\\
\varphi_{2}(x)\\
\end{array} \right)=0,
	\label{E2}
\end{eqnarray}
where we have introduced  $\varphi_{1}(x)$ and 
$\varphi_{2}(x)$ which are decomposed into upper and
lower components of the two-componentwave function $\psi(x)$. Eq. (\ref{E2}) turns into the two following coupled differential equations,
\begin{eqnarray}
	\label{E3}
	\frac{d\varphi_{1}(x)}{dx}&=&-\left[E-V(x)+m(x)\right]\varphi_{2}(x),\\
	\frac{d\varphi_{2}(x)}{dx}&=&\left[E-V(x)-m(x)\right]\varphi_{1}(x).
	\label{E4}
\end{eqnarray}
In order to simplify the solutions of these above equations, we use a two auxiliary components $\phi(x)$ and $\chi(x)$ in terms
of $\varphi_{1}(x)$ and $\varphi_{2}(x)$ as in\cite{b33} : 
\begin{eqnarray}
	\label{E5}
	\phi(x)&=&\varphi_{1}(x)+i\varphi_{2}(x),\\
	\chi(x)&=&\varphi_{1}(x)-i\varphi_{2}(x).
	\label{E6}
\end{eqnarray}
which lead
\begin{eqnarray}
	\label{E7}
	\frac{d\phi(x)}{dx}&=&i[E-V(x)]\phi(x)-im(x)\chi(x),\\
	\frac{d\chi(x)}{dx}&=&-i[E-V(x)]\chi(x)+im(x)\phi(x).
	\label{E8}
\end{eqnarray}
Eliminating $\chi(x)$ in Eq. (\ref{E7}) and inserting into Eq. (\ref{E8}) and
following the similar procedure for $\phi(x)$, we obtain two
uncoupled second-order differential equations for $\phi(x)$ and
$\chi(x)$, respectively
\begin{eqnarray}
	\frac{d^2\phi(x)}{dx^2}&-&\frac{dm(x)/dx}{m(x)}\frac{d\phi(x)}{dx}+\bigg\{[E-V(x)]^2-m^2(x)+i\frac{dV(x)}{dx},
	\nonumber\\&+&i[E-V(x)]\frac{dm(x)/dx}{m(x)}\bigg\}\phi(x)=0.
	\label{E9}
\end{eqnarray}
and
\begin{eqnarray}
	\frac{d^2\chi(x)}{dx^2}&-&\frac{dm(x)/dx}{m(x)}\frac{d\chi(x)}{dx}+\bigg\{[E-V(x)]^2-m^2(x)-i\frac{dV(x)}{dx}
	\nonumber\\&-&i[E-V(x)]\frac{dm(x)/dx}{m(x)}\bigg\}\chi(x)=0.
	\label{E10}
\end{eqnarray}
Now, we assume that the  Dirac particle mass distribution depends only on spatially coordinate and obeying a  cusp mass  form
\begin{eqnarray}
m(x)=m_{0}+m_{1}\left[\theta(-x)e^{\frac{x}{a}}+\theta(x)e^{-\frac{x}{a}}\right].\,
	\label{E11}
\end{eqnarray}
Here, the parameter $m_{0}$ will correspond to the rest mass of the Dirac particle
and $m_{1}$ is a real, positive, small parameter. In this regard, The mass distribution form supplies us to obtain the analytical results for the reflection and transmission coefficients and also the bound state solutions of the problematic. Besides, from Eq. (\ref{E9}), it is easy to see that the ratio of the derivative of the mass to
the mass is proportional with the mass parameter $m_{1}$. So, in order to simplify the solutions, we
rid the terms that contain the derivative of the mass in Eqs.
(\ref{E9}) and (\ref{E10}) for the case of $m_{1} \rightarrow 0$ \cite{b34}. So, Eqs. (\ref{E9}) and (\ref{E10}) becomes
\begin{equation}
\frac{d^2\phi(x)}{dx^2}+\bigg\{[E-V(x)]^2-m^2(x)+i\frac{dV(x)}{dx}\bigg\}\phi(x)=0,
	\label{E12}
\end{equation}
\begin{equation}
\frac{d^2\chi(x)}{dx^2}+\bigg\{[E-V(x)]^2-m^2(x)-i\frac{dV(x)}{dx}\bigg\}\chi(x)=0.
	\label{E13}
\end{equation}
Here, we consider the cusp potential possessing the form
\begin{eqnarray}
V(x)=V_{0}\left[\theta(-x)e^{\frac{x}{a}}+\theta(x)e^{-\frac{x}{a}}\right].\
	\label{E14}
\end{eqnarray}
The parameter $V_0$ represents the height of the barrier or the depth of the well depending on its sign. $a$ is a real, positive, parameter. Whereas $\theta(x)$ is the Heaviside step function.\\
In Fig. (\ref{fig1}), we display the shape form of the cusp potential for different values of the parameter a. 
\section{SCATTERING STATE SOLUTIONS}
In order to find the scattering of a Dirac particle under the cusp potential in the presence of the effective mass, we first seek the solution of the two uncoupled second-order differential equations (\ref{E9} and \ref{E10})  in the negative region ($x < 0$)  and the positive region ($x>0$). On the other hand, it is distinctly seen from (\ref{E7}) and (\ref{E8}) that it will be sufficient to solve one of (\ref{E12}) and (\ref{E13}) to obtain the solutions describing
the scattering states. These will be discussed in the subsections below.
\subsection{Solutions in the negative region: $x<0$}
By inserting the mass distribution form (\ref{E11}) and the cusp potential (\ref{E14}) into Eq. (\ref{E12}) and changing the variable $y=e^{\frac{x}{a}}$, we have
\begin{equation}
\frac{d^2\phi(y)}{dy^2}+\frac{1}{y}\frac{d\phi(y)}{dy}+\left[a^2\left(V_0^2-m_1^2\right)+\frac{\left(a\left(iV_0-2a\left(EV_0+m_0m_1\right)\right)\right)}{y}+\frac{a^2\left(E^2-m_0^2\right)}{y^2}\right]\phi(y)=0.
	\label{E15}
\end{equation}
By setting $\phi(y)=y^{-\frac{1}{2}}f(y)$ and substituting it into the above equation, one gets
the Whittaker differential equation possessing the form \cite{b35} :
\begin{equation}
  \frac{d^2f(z)}{dz^2}+\left[-\frac{1}{4}+\frac{\mu_{1L}}{z}+\frac{\left(\frac{1}{4}-K_{1L}^2\right)}{z^2}\right]f(z)=0,
  	\label{E16}
\end{equation}
where we have used the following parametrization 
\begin{equation}
\begin{cases}
& z=2a\sqrt{m_1^2-V_0^2}y=\beta_{1}y,\\
&\mu_{1L}=\frac{\left(i-2aE\right)V_0-2am_0m_1}{2\sqrt{m_1^2-V_0^2}},\\
& K_{1L}=a\sqrt{m_0^2-E^2}.
	\label{E17}
\end{cases}
\end{equation}
 Therefore the equation (\ref{E15}) has a general solution :
 \begin{equation}
 \phi_L(y)=A_1y^{-\frac{1}{2}}\mathcal{M}_{\mu_{1L},K_{1L}}\left(\frac{y}{\beta_{1}}\right)+A_2y^{-\frac{1}{2}}\mathcal{W}_{\mu_{1L},K_{1L}}\left(\frac{y}{\beta_{1}}\right).
 	\label{E18}
 \end{equation}
where $\mathcal{M}_{\mu,k}(z)$ and  $\mathcal{W}_{\mu,k}(z)$ are  Whittaker functions defined in terms of Kummer's confluent hypergeometric functions $M$ and $U$ by \cite{b35} :
\begin{eqnarray}
	\label{E19}
\mathcal{M}_{\mu,k}(z)=z^{k+\frac{1}{2}}e^{-\frac{z}{2}}M\left(k-\mu+\frac{1}{2},1+2k;z\right),\\
\mathcal{W}_{\mu,k}(z)=z^{k+\frac{1}{2}}e^{-\frac{z}{2}}U\left(k-\mu+\frac{1}{2},1+2k;z\right).
	\label{E20}
\end{eqnarray}
 \subsection{Solutions in the positive region: $x>0$}
 In order to find the scattering solutions for the other extreme $ x > 0$, we chose a new variable $y=e^{-\frac{x}{a}}$. So, by using, like Eq. (\ref{E15}), the transformation $\phi(y)=y^{-\frac{1}{2}}g(y)$ we also obtain the Whittaker differential equation
 \begin{equation}
 \frac{d^2g(t)}{dt^2}+\left[-\frac{1}{4}+\frac{\mu_{1R}}{t}+\frac{\left(\frac{1}{4}-K_{1R}^2\right)}{t^2}\right]g(t)=0,
 	\label{E21}
 \end{equation}
 where the result for the parameters $t$, $\mu_{1R}$ and $K_{1R}$  is found to be
 \begin{equation}
 \begin{cases}
 & t=2a\sqrt{m_1^2-V_0^2}y=\beta_{2}y,\\
 &\mu_{1R}=-\frac{\left(i+2aE\right)V_0+2am_0m_1}{2\sqrt{m_1^2-V_0^2}},\\
 & K_{1R}=a\sqrt{m_0^2-E^2}.
 	\label{E22}
 \end{cases}
 \end{equation}
 Thus, the equation (\ref{E21}), in this instance, has a general solution:
 \begin{equation}
 \phi_R(y)=A_3y^{-\frac{1}{2}}\mathcal{M}_{\mu_{1R},K_{1R}}\left(\frac{y}{\beta_{2}}\right)+A_4y^{-\frac{1}{2}}\mathcal{W}_{\mu_{1R},K_{1R}}\left(\frac{y}{\beta_{2}}\right).
 	\label{E23}
 \end{equation}
 So far we have derived the  analytical expressions  for the wave function of the scattering problem in the negative region $x < 0$ and in the positive region $x > 0$  which depend on four unknown constants ($A_1$, $A_2$) and ($A_3$, $A_4$) respectively  .
\subsection{Reflection and transmission coefficients}
In order to investigate the transmission and reflection coefficients, we should first check asymptotic behaviors of the results as $\pm\infty$.
So, for $x<0$  the asymptotic behavior of the wave function $\phi_L(x)$ when $x\rightarrow+\infty$ is
\begin{equation}
\phi_L(x\rightarrow-\infty)=A_1e^{iK_{1L}x}+A_2e^{-iK_{1L}x}.
	\label{E24}
\end{equation}
For the other extreme $x > 0$, we have
\begin{equation}
\phi_R(x\rightarrow+\infty)=A_3e^{iK_{1R}x}.
	\label{E25}
\end{equation}
It should be noticed, here, that $A_2=0$ because we have assumed the plane wave coming from the left to the right.
Going further and in order to get the electrical current density for the
one-dimensional Dirac equation defined by
\begin{eqnarray}
j=\frac{1}{2}\big[\left|\phi(x)\right|^2-\left|\chi(x)\right|^2\big],
	\label{E26}
\end{eqnarray}
we need to insert Eqs. (\ref{E24}) and (\ref{E25}) into Eq. (\ref{E7}) which gives
$\chi_{L}(x)$ and $\chi_{R}(x)$, respectively,
\begin{eqnarray}
	\label{E27}
\chi_{L}(x)&=&\frac{1}{m(x)}\left[(E-K_{1L})A_{1}e^{iK_{1L}x}+(E+K_{1L})A_{2}e^{-iK_{1L}x}\right],\\
\chi_{R}(x)&=&\left(\frac{E-K_{1R}}{m(x)}\right)A_{4}e^{iK_{1R}x}.
	\label{E28}
\end{eqnarray}
The current in Eq. (\ref{E26}) can be written as $j_{L}=j_{inc}+j_{refl}$
in the limit $x \rightarrow -\infty$ where $j_{inc}$ is the
incident and $j_{refl}$ is the reflected current. Similarly as $x
\rightarrow \infty$ the current is $j_{R}=j_{trans}$ where
$j_{trans}$ is the transmitted current. Inserting Eqs. (\ref{E24}), (\ref{E25}),
(\ref{E27}) and (\ref{E28}) into Eq. (\ref{E26}), we find the reflection and
transmission coefficients, respectively, as
\begin{eqnarray}
	\label{E29}
& R=\frac{j_{refl}}{j_{inc}}=&\frac{\left(E+K_{1L}\right)}{\left(E-K_{1L}\right)}\frac{|A_{2}|^2}{|A_{1}|^2}\,,\\
& T=\frac{j_{trans}}{j_{inc}}=&\frac{|A_{4}|^2}{|A_{1}|^2}\,.
	\label{E30}
\end{eqnarray}
Let’s use the following continuity conditions on the wave functions and their first derivatives
at $x=0$ to obtain explicit expressions for $R$ and $T$ :
 \begin{equation}
 \begin{cases}
 & \phi_L(x=0)=\phi_R(x=0),\\
 & \frac{d\phi_L(x)}{dx}|_{x=0}=\frac{d\phi_R(x)}{dx}|_{x=0}.
 \end{cases}
 	\label{E31}
 \end{equation}
 Using these equation with Eqs. (\ref{E29}) and (\ref{E30}), one can obtain
  \begin{align}
  	\label{E32}
  & A_1F_1+A_2F_2=A_4F_4,\\
  & A_1H_1+A_2H_2=A_4H_4,
  	\label{E33}
  \end{align}
  where the parameters $F_1$, $F_2$, $F_4$ and  $H_1$, $H_2$, $H_4$  are expressed as follows :
  \begin{equation}
 \begin{cases}
 	& F_1=\mathcal{M}_{\mu_{1L},K_{1L}}\left(\frac{1}{\beta_{1}}\right),\\
 	& F_2=\mathcal{W}_{\mu_{1L},K_{1L}}\left(\frac{1}{\beta_{1}}\right),\\
 	& F_4=\mathcal{W}_{\mu_{1R},K_{1R}}\left(\frac{1}{\beta_{2}}\right),\\ 
 	 & H_1=\frac{\left(1-\left(1+2\mu_{1L}\right)\beta_1\right)\mathcal{M}_{\mu_{1L},K_{1L}}\left(\frac{1}{\beta_{1}}\right)+\left(1+2\left(\mu_{1L}+K_{1L}\right)\right)\beta_1\mathcal{M}_{\mu_{1L}+1,K_{1L}}\left(\frac{1}{\beta_{1}}\right)}{2a\beta_1},\\
 	 & H_2=\frac{\left(1-\left(1+2\mu_{1L}\right)\beta_1\right)\mathcal{W}_{\mu_{1L},K_{1L}}\left(\frac{1}{\beta_{1}}\right)-2\beta_1\mathcal{W}_{\mu_{1L}+1,K_{1L}}\left(\frac{1}{\beta_{1}}\right)}{2a\beta_1},\\
 	 & H_4=\frac{2\beta_2\mathcal{W}_{\mu_{1R}+1,K_{1R}}\left(\frac{1}{\beta_{2}}\right)+\left(\left(2\mu_{1R}+1\right)\beta_2-1\right)\mathcal{W}_{\mu_{1R},K_{1R}}\left(\frac{1}{\beta_{2}}\right)}{2a\beta_2}.
 \end{cases}
 	\label{E34}
\end{equation}
Therefore, transmission and reflection coefficients are, respectively, obtained as follows :
\begin{eqnarray}
	\label{E35}
& R=&\frac{\left(E+K_{1L}\right)}{\left(E-K_{1L}\right)}\frac{|F_4H_1-F_1H_4|^2}{|F_2H_4-F_4H_2|^2}\,,\\
& T=&\frac{|F_2H_1-F_1H_2|^2}{|F_2H_4-F_4H_2|^2}\,.
	\label{E36}
\end{eqnarray}
\section{Results and Discussion}
\begin{figure}[H] 
	\centering
	\rotatebox{0}{\includegraphics[height=43mm]{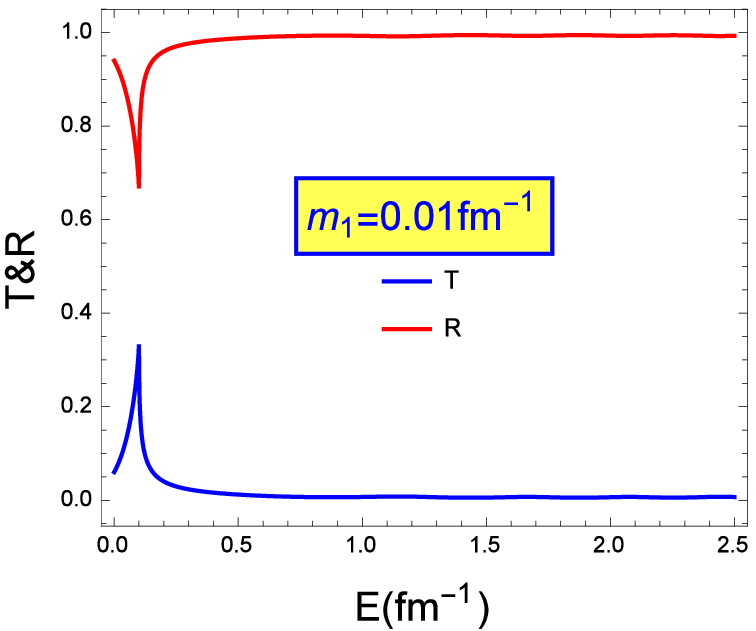}} 
	\rotatebox{0}{\includegraphics[height=43mm]{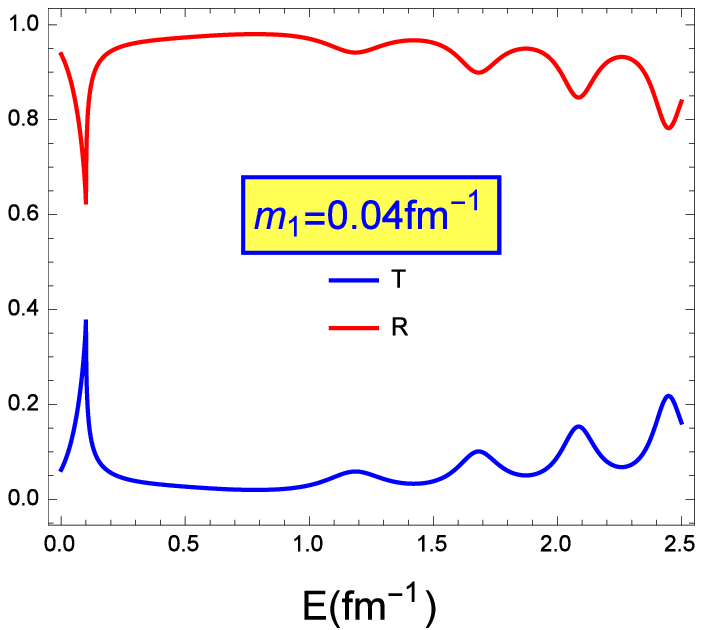}}
	\rotatebox{0}{\includegraphics[height=43mm]{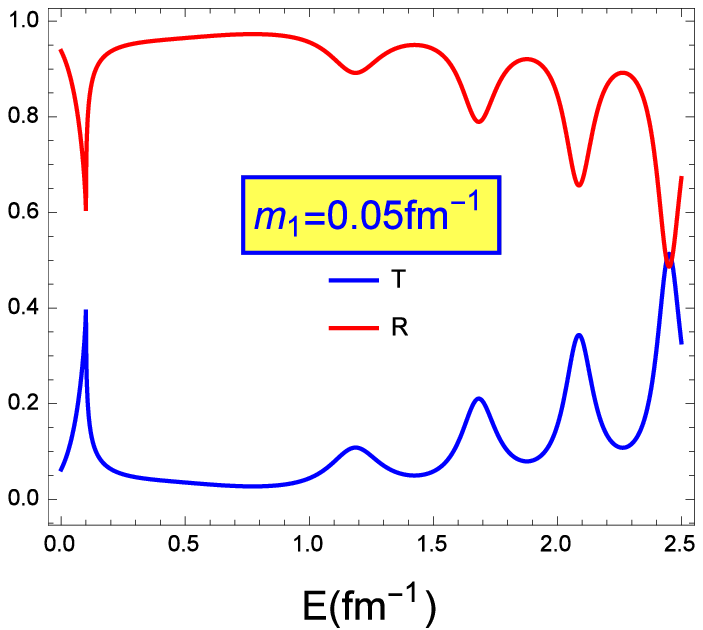}}
	\caption{Transmission and reflection coefficients as a function of the energy E for $m_1=0.01fm^{-1}$(left), $m_1=0.04fm^{-1}$ (middle) and $m_1=0.05fm^{-1}$ (right).
		The others parameters are taking as $m_0=0.1fm^{-1}$, $V_0=0.4fm^{-1}$ and $a=2.5fm$.}
	\label{fig2}
\end{figure}

\begin{figure}[H]
	\centering
	\rotatebox{0}{\includegraphics[height=43mm]{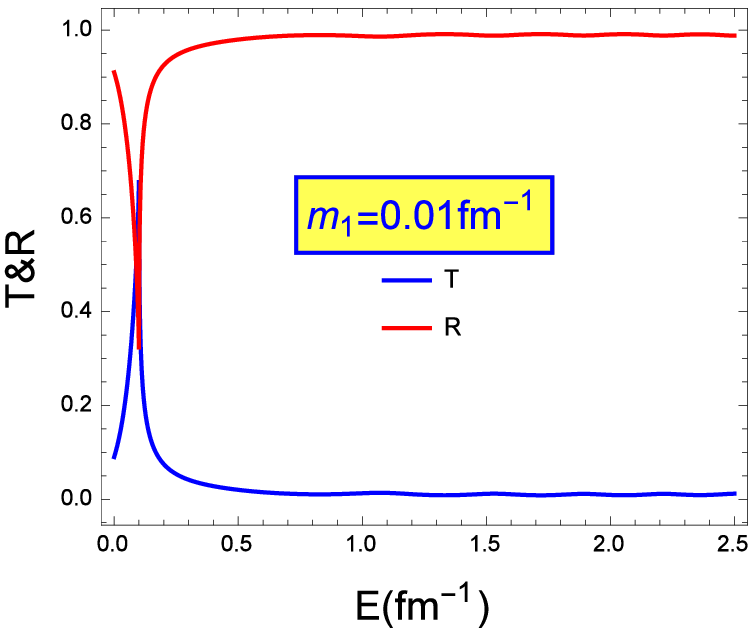}} 
	\rotatebox{0}{\includegraphics[height=43mm]{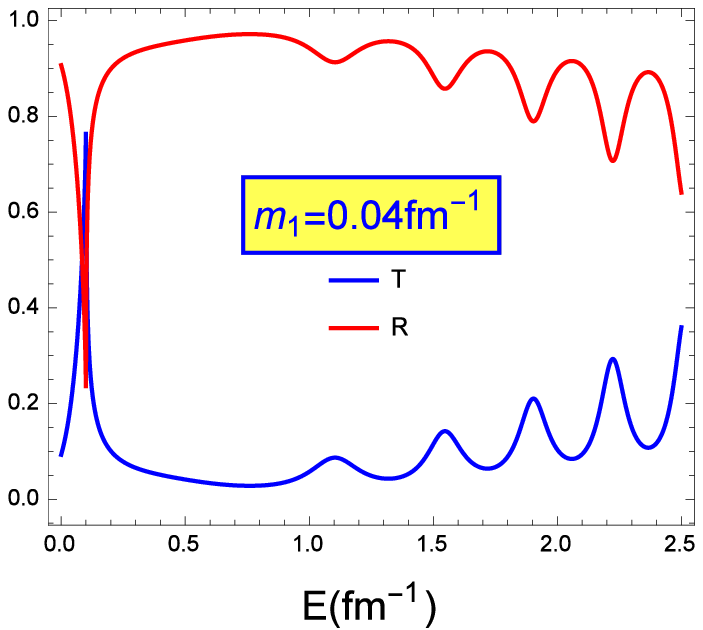}}
	\rotatebox{0}{\includegraphics[height=43mm]{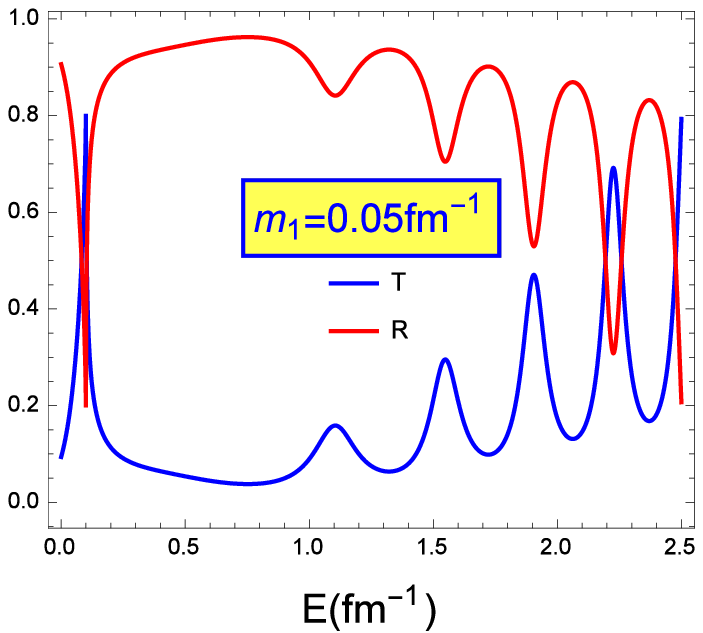}}
	\caption{The same as in Fig. \ref{fig2}, but for $a=3fm$.}
	\label{fig3}
\end{figure}

\begin{figure}[H]
	\centering
	\rotatebox{0}{\includegraphics[height=43mm]{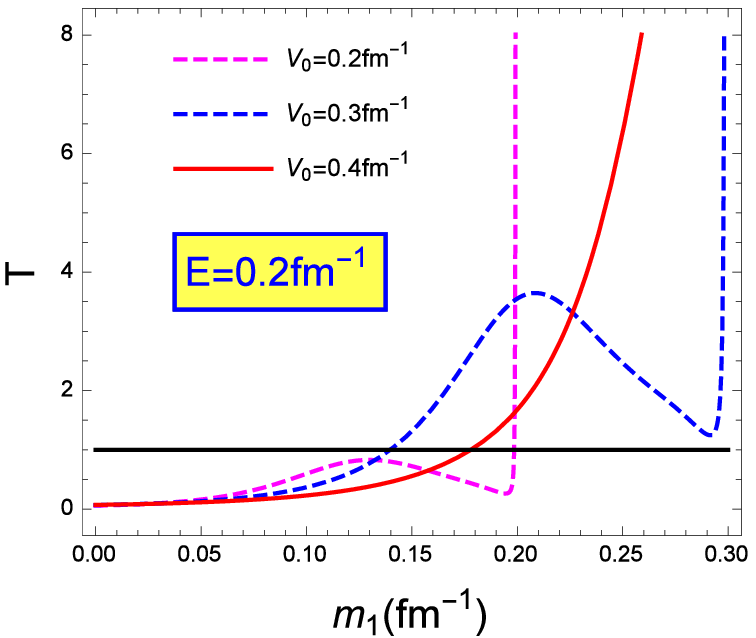}} 
	\rotatebox{0}{\includegraphics[height=43mm]{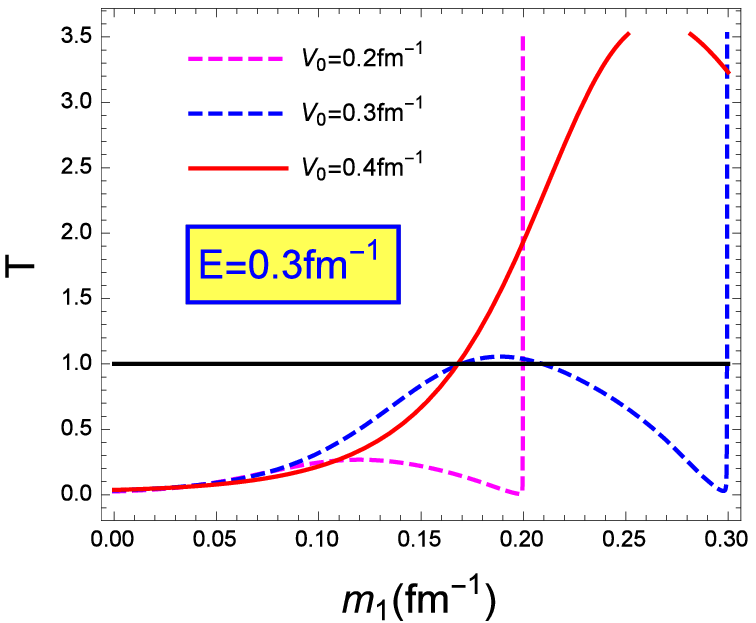}}
	\rotatebox{0}{\includegraphics[height=43mm]{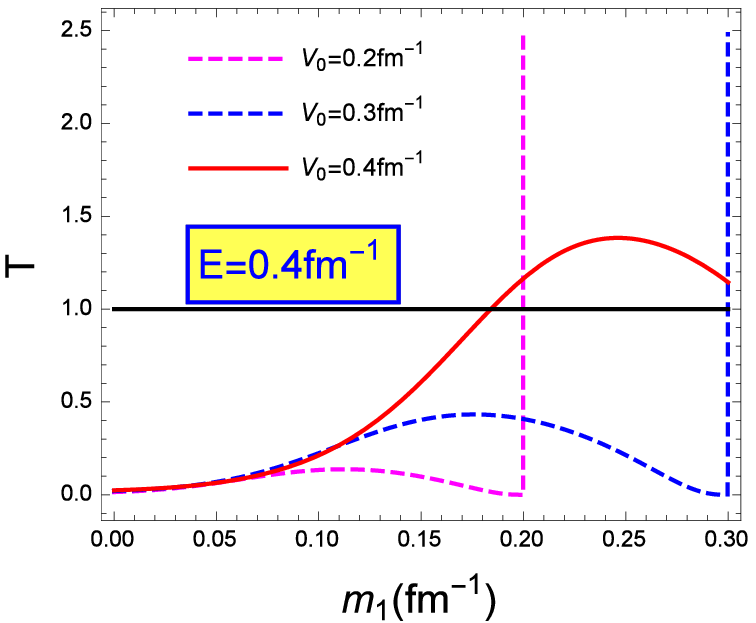}}
	\caption{Coefficient of transmission versus the mass parameter $m_1$ with  several values of $V_0$. Left plots are for
		($E=0.2fm^{-1},a=3fm,m_0=0.1fm^{-1}$), middle plots for	($E=0.3fm^{-1},a=3fm,m_0=0.1fm^{-1}$) and  right plots for	($E=0.4fm^{-1},a=3fm,m_0=0.1fm^{-1}$). }
	\label{fig4}
\end{figure}

In this section, we present  the main results of our theoretical study in which  the effect of various parameters on  the reflection and transmission coefficients of the cusp potential barrier are appropriately discussed. So, in Fig. (\ref{fig2}), the transmission and reflection coefficients are displayed as function of the energy $E$ for different values of the mass parameter  $m_1$. From this Figure, one can see that the transmission and reflection coefficients oscillate between the values zero and one and satisfy the condition $ R+T=1$. 
Also, from figures (\ref{fig2}) and (\ref{fig3}), one can see that for a lower value of the mass parameter $m_1$, namely: $m_1=0.01fm^{-1}$, there is only one resonance peak at the energy $E=m_0=0.01fm^{-1}$. Such a peak corresponding to transmission threshold energy appears also in spectra for relatively higher values of $m_1$. But, as $m_1$ increases, more resonance peaks start to appear at higher energies. The most pronounced one of them occurs at the energy $E=2.45fm^{-1}$ for $m_1=0.05fm^{-1}$. The occurrence of several resonance peaks at slightly higher values of the parameter $m_1$ shows that the potential barrier becomes more transparent with the increase of $m_1$. However, we have to notice that the mass parameter $m_1$ cannot increase so much otherwise the reflection and transmission coefficients will diverge due to the limit condition $m_1\rightarrow0$ being set above  for solution of the present scattering problem. Such a fact is well illustrated in Fig. (\ref{fig4}) for transmission coefficient . From this figure, one can observe that the behavior of transmission coefficient $T$ depends also on the potential barrier height. Besides, the length parameter a has a significant effect only on the amplitude of the reflection and transmission coefficients as can be seen from figures (\ref{fig2}) and (\ref{fig3}). While resonance peak position is slightly affected. The increase of such a parameter leads to an overlap of both reflection and transmission peaks at resonances.
\section{Conclusion}
We studied the scattering states of Dirac equation in the presence of the cusp potential, considering position-dependency for mass, in one dimension in an analytical manner. We first obtained complicated coupled differential equation governing on the wave function components. Introducing a two auxiliary components caused to have a simpler differential equations for the components whose solutions were written in terms of Kummers confluent hypergeometric functions. Since the cusp potential has a delta-like essence, we had to derive reflection and transmission coefficients. After deriving the reflection and transmission coefficients, we plotted the results in terms of energy and the real mass parameter. It was shown that the increase of this latter leads to the appearance of several resonance peaks beyond the typical one corresponding to the transmission threshold energy $E=m_0$ . So, the potential barrier becomes more transparent. Also, it was shown that the length parameter a has an important impact on the amplitude of both reflection and transmission coefficients leading to an overlap of their corresponding peaks at resonances conserving the condition $(T+R=1)$. The effect of the potential barrier heigth on the behavior of the transmission coefficient was also illustrated.

\section{Acknowledgement}
The authors wish to thank the referees for their valuable suggestions and recommendations which helped to improve the paper.

\end{document}